\documentclass[twocolumn,amsmath,showpacs,amsfonts,aps,prc,superscriptaddress]{revtex4} 

\usepackage{epsfig}
\usepackage{bm}

\newcommand{\2}{\frac{1}{2}}
\newcommand{\half}{{\textstyle\2}}

\newcommand\bea{\begin{eqnarray}}
\newcommand\eea{\end{eqnarray}}

\newcommand{\bb}{\bm{b}}
\newcommand{\bq}{\bm{q}}
\newcommand{\bK}{\bm{K}}

\newcommand{\s}{{\cal S}}
\newcommand{\C}{{\cal C}}

\newcommand{\Eq}{{\,=\,}}
\newcommand{\Kt}{K_\perp}

\begin{document}


\title{Symmetry constraints for the emission angle dependence\\
 of Hanbury-Brown-Twiss radii}

\author{U. Heinz}
\affiliation{Physics Department, The Ohio State University, Columbus,
 OH 43210}
\affiliation{Kavli Institute for Theoretical Physics, University of
California, Santa Barbara, CA 93106-4030}
\author{A. Hummel}
\affiliation{Physics Department, The Ohio State University, Columbus,
 OH 43210}
\author{M.A. Lisa}
\affiliation{Physics Department, The Ohio State University, Columbus,
 OH 43210}
\author{U.A. Wiedemann}
\affiliation{Theoretical Physics Division, CERN, CH-1211 Geneva 23,
 Switzerland}

\begin{abstract} 
We discuss symmetry constraints on the azimuthal oscillations of
two-particle correlation (Hanbury Brown--Twiss interferometry) radii
for non-central collisions between equal spherical nuclei. We also 
propose a new method for correcting in a model-independent way the 
emission angle dependent correlation function for finite event plane 
resolution and angular binning effects. 
\end{abstract}


\preprint{NSF-ITP-02-48}

\date{July 15, 2002\ (revised version)}  

\maketitle

\section{Introduction} 
\label{sec1}  

In the study of heavy ion collisions two-particle Bose-Einstein 
correlations are an important tool for extracting information on 
the space-time structure of the collision zone at freeze-out \cite{WH99}.
At non-zero impact parameter the reaction zone formed by the two 
overlapping nuclei is initially spatially deformed. When 
viewed along the beam direction $z$, it is longer in the direction 
$y$ perpendicular to the reaction plane (defined by the beam axis and 
the impact parameter vector pointing in $x$ direction) than in the 
reaction plane. The reaction plane orientation can be determined
event-by-event from anisotropies in the collective flow of the
emitted particles \cite{O93,VZ96,PV98}: at lower collision energies 
one exploits the {\em directed flow} of the protons near projectile 
and target rapidity (``bounce off'') \cite{SG86}, while at high energies 
(where the directed flow becomes very weak) one uses the {\em elliptic 
flow} of the produced particles at midrapidity \cite{O92}.

Having extracted the orientation of the reaction plane from the
final distribution of the emitted particle {\em momenta},
one can then address the question of their {\em spatial} distribution 
relative to the reaction plane by measuring \cite{E895,R02} two-particle 
correlations as a function of the azimuthal emission angle $\Phi$ 
(i.e. the direction of the transverse momentum vector $\bK_\perp$ of 
the emitted particle pairs relative to the impact parameter $\bb$) 
\cite{VC95,W97,LHW00}. Complementing the spectral information on the
momentum-space structure of the source with space-time information 
from the correlation functions severely constrains models for the 
dynamical evolution of the reaction zone \cite{WH99}. For noncentral
collisions interesting questions which can be addressed in this way 
are the origin and manifestation of anisotropic collective flow and 
its consequences for the space-time evolution of the fireball, from 
which information about the intensity of rescattering effects and the 
degree of thermalization in particular during the early stages of the 
collision can be extracted (see, e.g., \cite{HK02}).   

For quantum statistical correlations between, say, identical pions 
the measured two-pion correlation function $C(\bq,\bK)$ is related 
to the emission function (single-pion phase-space distribution 
at freeze-out) $S(x,\bK)$ by \cite{WH99}
 \begin{equation}
 \label{1}
   C(\bm{q},\bm{K}) = 1 + \left| \left\langle 
                    e^{i\bm{q}{\cdot}(\bm{x}-\bm{\beta}t)}
                    \right\rangle(\bm{K})\right|^2  .
 \end{equation}
Here $\bm{q}\Eq\bm{p}_1{-}\bm{p}_2$ and $\bK\Eq\2(\bm{p}_1{+}\bm{p}_2)$ 
are the relative and average momentum of the pion pair, respectively, 
$\bm{\beta}\Eq(\bm{p}_1{+}\bm{p}_2)/(E_1{+}E_2)$ is the velocity of the
pair, and the average $\langle\dots\rangle$ is taken with the emission 
function:
 \begin{equation}
 \label{2}
  \langle f(x)\rangle(\bK) = \frac{\int d^4x\, f(x)\,S(x,\bK)}
                                  {\int d^4x\, S(x,\bK)}\,.
 \end{equation}

If the space-time structure of $S(x,\bK)$ can be approximated by a 
Gaussian, the resulting correlator $C(\bq,\bK)$ is again a Gaussian 
in the relative momentum $\bq$ and can be fully characterized by 
six {\em HBT radius parameters} $R^2_{ij}$ which are functions 
of the pair momentum $\bK$ \cite{WH99}:
 \begin{equation}
 \label{3}
  C(\bq,\bK)= 1 + \exp\biggl[-\sum_{i,j{=}o,s,l} q_i q_j R^2_{ij}(\bK)\biggr]
  \, .
 \end{equation}
Here $\bq$ is conventionally decomposed into orthogonal components along
the beam direction ($l\Eq{longitudinal}$), parallel to the transverse
pair momentum $\bK_\perp$ ($o\Eq{out}$) and along the remaining third 
direction ($s\Eq{side}$). In this $(osl)$-frame the pair velocity has 
components $\bm{\beta}=(\beta_\perp,0,\beta_l)$. The radius parameters 
$R_{ij}^2$ are then calculable from the spatial correlation tensor 
 \begin{equation}
 \label{4}
  S_{\mu\nu}(\bK)=\langle x_\mu x_\nu\rangle(\bK)
  - \langle x_\mu\rangle(\bK) \langle x_\nu\rangle(\bK)
  \equiv \langle\tilde x_\mu\tilde x_\nu\rangle,
 \end{equation}
$(\mu,\nu\Eq0,1,2,3)$, 
which describes, for pairs with momentum $\bK$, the widths in space-time 
of the emission function $S(x,\bK)$ around the point of highest 
emissivity \cite{WH99}. The spatial correlation tensor is specified in 
coordinates $x_\mu$ attached to the reaction plane: $x_3\Eq{z}$ is the 
beam direction, $x_1\Eq{x}$ is the direction of the impact parameter 
$\bb$, and $x_2\Eq{y}$ points perpendicular to the reaction plane. For 
the spatial correlation tensor this choice of coordinates is natural 
since the reaction plane is a symmetry plane for the collision. The 
relations between $R^2_{ij}$ and $S_{\mu\nu}$ are \cite{W97,WH99}
  \bea
  \label{5}
    R_s^2 &=& \textstyle{\2}(S_{11}{+}S_{22}) 
            - \textstyle{\2}(S_{11}{-}S_{22})\cos(2\Phi)
            - S_{12} \sin(2\Phi)
  \nonumber\\
    R_o^2 &=& \textstyle{\2}(S_{11}{+}S_{22}) 
            + \textstyle{\2}(S_{11}{-}S_{22})\cos(2\Phi)
            + S_{12} \sin(2\Phi)
  \nonumber\\*
          &&- 2\beta_\perp (S_{01} \cos\Phi{+}S_{02} \sin\Phi)
             + \beta_\perp^2 S_{00}, 
  \nonumber\\
    R_{os}^2 &=& S_{12} \cos(2\Phi) 
        - \textstyle{\2} \left(S_{11}{-}S_{22}\right)\sin(2\Phi)
  \nonumber\\*
             &&+ \beta_\perp (S_{01} \sin\Phi{-}S_{02} \cos\Phi), 
  \nonumber\\
    R_{l}^2 &=& S_{33} -2 \beta_l S_{03} + \beta_l^2 S_{00}, 
  \nonumber\\
    R_{ol}^2 &=& \left( S_{13}{-}\beta_l S_{01}\right) \cos\Phi
               + \left( S_{23}{-}\beta_l S_{02}\right) \sin\Phi
  \nonumber\\
             &&- \beta_\perp S_{03} 
               + \beta_l\beta_\perp S_{00},
  \nonumber\\
    R_{sl}^2 &=& \left(S_{23}{-}\beta_l S_{02}\right) \cos\Phi
               - \left(S_{13}{-}\beta_l S_{01}\right) \sin\Phi .
  \eea

The radius parameters $R_{ij}^2$ are functions of the pair rapidity 
$Y=\2\ln[(1{+}\beta_l)/(1{-}\beta_l)]$, the magnitude $\Kt$ of 
the transverse pair momentum, and its angle $\Phi$ relative to
the reaction plane (the azimuthal emission angle). In Eqs.~(\ref{5}) 
we only indicated the {\em explicit} $\Phi$ dependence arising from 
the azimuthal rotation of the $(osl)$ system relative to the 
reaction-plane-fixed $(xyz)$ system:
 \begin{equation}
 \label{6}
   x_o= x\cos\Phi+y\sin\Phi,\quad x_s=-x\sin\Phi+y\cos\Phi.
 \end{equation}
In addition to this explicit $\Phi$ dependence, there is an 
{\em implicit} one \cite{W97} arising from the dependence of the 
emission function $S(x,\bK){\Eq}S(x,y,z,t;Y,\Kt,\Phi)$ on the emission 
angle $\Phi$; this generates a $\Phi$ dependence of the components 
of the spatial correlation tensor $S_{\mu\nu}$. In this note 
we work out the symmetry constraints on the $\Phi$ dependence of 
$S_{\mu\nu}$ and study their implications on the $\Phi$ dependence
of the HBT radius parameters after the explicit $\Phi$ dependence shown
in Eqs.~(\ref{5}) is folded in.

\section{Symmetries of the emission function and spatial correlation 
            tensor}
\label{sec2}

For spherical colliding nuclei the emission function is symmetric under 
reflection at the reaction plane:
 \begin{equation}
 \label{7}
   \text{I}:\, S(x,y,z,t;Y,\Kt,\Phi) = S(x,{-}y,z,t;Y,\Kt,-\Phi).
 \end{equation}
This leads to the following symmetry relations for the spatial correlation
tensor:
 \begin{equation}
 \label{8}
   \text{I}: S_{\mu\nu}(Y,\Kt,\Phi) = \theta_1 S_{\mu\nu}(Y,\Kt,-\Phi)\,,
 \end{equation}
with
 \begin{equation}
 \label{9}
   \theta_1 = (-1)^{\delta_{\mu 2}+\delta_{\nu 2}}.
 \end{equation}
Thus, symmetry I relates the components of $S_{\mu\nu}$ at emission angle
$\Phi$ with those at angle $-\Phi$ at the same pair rapidity $Y$ and 
transverse momentum $K_\perp$. $S_{02},\, S_{12}$ and $S_{23}$ are odd
under this symmetry ($\theta_1\Eq{-}1$), all other components are
even ($\theta_1\Eq{+}1$).  

If the two nuclei have equal mass, the emission function is also symmetric
under interchange of projectile and target. In the center of mass system
centered at the collision point, this translates into a point reflection 
symmetry at the origin:
 \bea
 \label{10}
   \text{II}: && S(x,y,z,t;Y,\Kt,\Phi) 
 \nonumber\\
   &&\qquad\quad
     = S(-x,-y,-z,t;-Y,\Kt,\Phi+\pi).\quad
 \eea
For the spatial correlation tensor this implies
 \begin{equation}
 \label{11} 
   \text{II}: S_{\mu\nu}(Y,\Kt,\Phi) = \theta_2\,
   S_{\mu\nu}(-Y,\Kt,\Phi{+}\pi),
 \end{equation}
with
 \begin{equation}
 \label{12}
   \theta_2 = (-1)^{\delta_{\mu 0}+\delta_{\nu 0}}.
 \end{equation}
Symmetry II relates $S_{\mu\nu}$ at emission angle $\Phi$ for forward-going 
pairs ($Y>0$) with $S_{\mu\nu}$ at emission angle $\Phi+\pi$ for 
backward-going pairs ($Y<0$), and vice versa. For midrapidity pairs 
($Y\Eq0$) it relates the spatial correlation tensor at emission angles 
$\Phi$ and $\Phi{+}\pi$, providing a useful second constraint on the
emission angle dependence. $S_{01},\,S_{02}$ and $S_{03}$ are odd under 
this symmetry ($\theta_2\Eq{-}1$) while all other components of 
$S_{\mu\nu}$ are even ($\theta_2\Eq{+}1$). 

Finally, at very high collision energies the source is expected to
be approximately invariant under longitudinal boosts within an extended
rapidity interval around $Y\Eq0$. If this is the case, the emission 
function $S(x,\bK)$, when expressed in terms of longitudinal proper
time $\tau\Eq\sqrt{t^2{-}z^2}$ and space-time rapidity 
$\eta\Eq\2\ln[(t{+}z)/(t{-}z)]$, depends only on the difference 
$\eta{-}Y$ between the space-time and momentum-space rapidities. For 
equal projectile and target nuclei it then must be an even function
of $\eta{-}Y$, i.e. invariant under a simultaneous sign change of $Y$
and $\eta$. With $z\Eq\tau\sinh\eta$ and $t\Eq\tau\cosh\eta$ this
implies
 \begin{equation}
 \label{13}
   \text{III}: S(x,y,z,t;Y,\Kt,\Phi) = S(x,y,-z,t;-Y,\Kt,\Phi)
 \end{equation}
and
 \begin{equation}
 \label{14} 
   \text{III}: S_{\mu\nu}(Y,\Kt,\Phi) = \theta_3\,
   S_{\mu\nu}(-Y,\Kt,\Phi).
 \end{equation}
with
 \begin{equation}
 \label{15}
   \theta_3 = (-1)^{\delta_{\mu 3}+\delta_{\nu 3}}.
 \end{equation}
Combining symmetries II and III allows to relate the spatial correlation 
tensor at angles $\Phi$ and $\Phi{+}\pi$ {\em for all rapidities $Y$}.
For boost-invariant sources, the terms with $\theta_3\Eq{-}1$ 
(i.e. $S_{03}$, $S_{13}$, and $S_{23}$) vanish at $Y\Eq0$. We note 
that the symmetry (\ref{13}) also applies to sources {\em without} 
boost-invariance if they exhibit spatial and momentum anisotropies
(i.e. $\Phi$-dependence) already at {\em zero} impact parameter, 
such as fully central collisions (no spectators) between deformed
nuclei (e.g. U+U). In this case the source is symmetric under the
simultaneous reflection of coordinates and momenta at the transverse
plane at $z\Eq0$, in agreement with Eq.~(\ref{13}) \cite{Kolb}.

We will concentrate here on the consequences of the combination of 
symmetries I and II at $Y\Eq0$ and of the combination of all three 
symmetries at any $Y$. The former case is relevant for two-pion 
correlations at midrapidity in low-energy collisions between equal
spherical nuclei, the latter case applies to high energy collisions,
such as those studied at the heavy-ion colliders RHIC and LHC. 
Symmetry I alone is less restrictive and is the only useful one 
when significantly away from midrapidity (in particular in the 
projectile and target fragmentation regions).

\section{Azimuthal Fourier decomposition of the spatial correlation tensor}
\label{sec3}

%
\begin{table}[t]
\caption{\label{T1} Consequences of symmetries I and II (see text) for
the azimuthal Fourier expansion of the spatial correlation tensor 
$S_{\mu\nu}$ at midrapidity $Y\Eq0$. The last column lists the angles
$\Phi$ in the first quadrant where $S_{\mu\nu}(Y\Eq0,\Kt,\Phi)$ 
vanishes. The notation follows the one introduced in \cite{W97}.}
\medskip 
\begin{ruledtabular}
\begin{tabular}{c|c|c|c|c} 
   \ $S_{\mu\nu}$\        & $\theta_1$ \  &
   $\theta_2$ \ &  Fourier expansion\  &
   Zeroes \ \\ \hline
  $\frac{\langle\tilde x^2{+}\tilde y^2\rangle}{2}$  &  1  &  1  & 
  $A_0 + 2\sum_{n\geq2,{\rm even}} A_n\cos(n\Phi)$ \ & --   \\ 
  $\frac{\langle\tilde x^2{-}\tilde y^2\rangle}{2}$  &  1  &  1  & 
  $B_0 + 2\sum_{n\geq2,{\rm even}} B_n\cos(n\Phi)$ \ & --   \\ 
  $\langle\tilde x\tilde y\rangle$  &  -1  &  1  & 
  $2\sum_{n\geq2,{\rm even}} C_n\sin(n\Phi)$ & $0^\circ,90^\circ$ \\ 
  $\langle\tilde t^2\rangle$  &  1  &  1  & 
  $D_0 + 2\sum_{n\geq2,{\rm even}} D_n\cos(n\Phi)$ \ & --   \\ 
  $\langle\tilde t\tilde x\rangle$  &  1  &  -1  & 
  $2\sum_{n\geq1,{\rm odd}} E_n\cos(n\Phi)$ & $90^\circ$ \\ 
  $\langle\tilde t\tilde y\rangle$  &  -1  &  -1  & 
  $2\sum_{n\geq1,{\rm odd}} F_n\sin(n\Phi)$ & $0^\circ$ \\ 
  $\langle\tilde t\tilde z\rangle$  &  1  &  -1  & 
  $2\sum_{n\geq1,{\rm odd}} G_n\cos(n\Phi)$ & $90^\circ$ \\ 
  $\langle\tilde x\tilde z\rangle$  &  1  &  1  & 
  $H_0 + 2\sum_{n\geq2,{\rm even}} H_n\cos(n\Phi)$ \ & --   \\ 
  $\langle\tilde y\tilde z\rangle$  &  -1  &  1  & 
  $2\sum_{n\geq2,{\rm even}} I_n\sin(n\Phi)$ & $0^\circ,90^\circ$ \\ 
  $\langle\tilde z^2\rangle$  &  1  &  1  & 
  $J_0 + 2\sum_{n\geq2,{\rm even}} J_n\cos(n\Phi)$ \ & --   \\ 
\end{tabular}
\end{ruledtabular}
\end{table} 
%
The above symmetries constrain the $\Phi$ dependence of the components
of the spatial correlation tensor (and thereby the {\em implicit} 
$\Phi$ dependence of the HBT radius parameters in Eq.~(\ref{5})). 
Correspondingly, certain expansion coefficients will vanish in an 
azimuthal Fourier expansion of $S_{\mu\nu}$. Let us write generically
$S(\Phi)$ for the $\Phi$-dependence of a given component $S_{\mu\nu}$.
Being a real function it has the following Fourier decomposition:
 \bea
 \label{16}
   S(\Phi) &=& \C_0 + 2\sum_{n=1}^\infty\left[\C_n\cos(n\Phi) 
                                            + \s_n\sin(n\Phi)\right],
 \\
 \label{17}
   \C_n &=& \int_{-\pi}^\pi \frac{d\Phi}{2\pi}\, S(\Phi)\,\cos(n\Phi)\,,
 \nonumber\\
   \s_n &=& \int_{-\pi}^\pi \frac{d\Phi}{2\pi}\, S(\Phi)\,\sin(n\Phi)\,.
 \eea
Symmetry I implies
 \bea
 \label{18}
   \theta_1 &=& +1 \quad\Longrightarrow\quad  \s_n=0\;\text{for\ all}\;n,
 \nonumber\\
   \theta_1 &=& -1 \quad\Longrightarrow\quad  \C_n=0\;\text{for\ all}\;n.
 \eea
At $Y\Eq0$, symmetry II eliminates even or odd terms in the Fourier series:
 \bea
 \label{19}
   \theta_2 &=& +1 \quad\Longrightarrow\quad  
   \C_n,\s_n=0\;\text{for\ odd}\;n, 
 \nonumber\\
   \theta_2 &=& -1 \quad\Longrightarrow\quad 
   \C_n,\s_n=0\;\text{for\ even}\;n.
 \eea     
%
\begin{table}
\caption{\label{T2} Consequences of symmetries I, II and III (see text) 
for the azimuthal Fourier expansion of the spatial correlation tensor 
$S_{\mu\nu}$ for boost-invariant sources. The last column lists the 
rapidities $Y$ and angles $\Phi$ in the first quadrant where 
$S_{\mu\nu}(Y,\Kt,\Phi)$ vanishes.}
\medskip 
\begin{ruledtabular}
\begin{tabular}{c|c|c|c|c} 
  \ $S_{\mu\nu}$  \  &~$\theta_3$~&~$\theta_2\theta_3$~&
  \ Fourier expansion \ & \ Zeroes \ \\ \hline
  $\frac{\langle\tilde x^2{+}\tilde y^2\rangle}{2}$ &  1  &  1  & 
  $A_0{+}2\sum_{n\geq2,{\rm even}} A_n\cos(n\Phi)$ & --   \\ 
  $\frac{\langle\tilde x^2{-}\tilde y^2\rangle}{2}$ &  1  &  1  & 
  $B_0{+}2\sum_{n\geq2,{\rm even}} B_n\cos(n\Phi)$ & --   \\ 
  $\langle\tilde x\tilde y\rangle$  &  1  &  1  & 
  $2\sum_{n\geq2,{\rm even}} C_n\sin(n\Phi)$ & $\Phi{=}0^\circ,90^\circ$ \\ 
  $\langle\tilde t^2\rangle$  &  1  &  1  & 
  $D_0{+}2\sum_{n\geq2,{\rm even}} D_n\cos(n\Phi)$ & --   \\ 
  $\langle\tilde t\tilde x\rangle$  &  1  &  -1  & 
  $2\sum_{n\geq1,{\rm odd}} E_n\cos(n\Phi)$ & $\Phi{=}90^\circ$ \\ 
  $\langle\tilde t\tilde y\rangle$  &  1  &  -1  & 
  $2\sum_{n\geq1,{\rm odd}} F_n\sin(n\Phi)$ & $\Phi{=}0^\circ$ \\ 
  $\langle\tilde t\tilde z\rangle$  &  -1  &  1  & 
  $G_0{+}2\sum_{n\geq2,{\rm even}} G_n\cos(n\Phi)$ & $Y{=}0$ \\ 
  $\langle\tilde x\tilde z\rangle$  &  -1  &  -1  & 
  $2\sum_{n\geq1,{\rm odd}} H_n\cos(n\Phi)$ & $Y{=}0,\Phi{=}90^\circ$ \\ 
  $\langle\tilde y\tilde z\rangle$  &  -1  &  -1  & 
  $2\sum_{n\geq1,{\rm odd}} I_n\sin(n\Phi)$ & $Y{=}0,\Phi{=}0^\circ$ \\ 
  $\langle\tilde z^2\rangle$  &  1  &  1  & 
  $J_0{+}2\sum_{n\geq2,{\rm even}} J_n\cos(n\Phi)$ & --   \\ 
\end{tabular}
\end{ruledtabular}
\end{table} 
%
In fact, Eq.~(\ref{11}) implies a stronger result:
 \bea
 \label{19a}
   \theta_2 = +1 &\Longrightarrow&
   \C_n,\s_n\;\text{are\ odd\ (even)\ functions\ of}\ Y
 \nonumber\\
   &&\ \text{for\ odd\ (even)\ values\ of}\;n, 
 \nonumber\\
   \theta_2 = -1 &\Longrightarrow&
   \C_n,\s_n\;\text{are\ odd\ (even)\ functions\ of}\ Y
 \nonumber\\
   &&\ \text{for\ even\ (odd)\ values\ of}\;n. 
 \eea     
This will be used in Section~\ref{sec5}. At $Y\Eq0$ Eq.~(\ref{19})
follows from (\ref{19a}).

Table~\ref{T1} lists the Fourier expansions for the components of the
spatial correlation tensor at midrapidity which result from the combination
of these two symmetries. Following \cite{W97} we have used the fact that
$S_{11}$ and $S_{22}$ have structurally identical Fourier expansions
and combined them into $A\Eq\2(S_{11}{+}S_{22})$ and 
$B\Eq\2(S_{11}{-}S_{22})$, which are the combinations entering in 
Eqs.~(\ref{5}). 

For boost-invariant sources, we can combine symmetries II and III
(Eqs.(\ref{10}) and (\ref{13})) to obtain
 \begin{equation}
 \label{20} 
   S_{\mu\nu}(Y,\Kt,\Phi) = \theta_2\theta_3\,S_{\mu\nu}(Y,\Kt,\Phi+\pi).
 \end{equation}
For the Fourier coefficients this implies
 \bea
 \label{21}
   \theta_2\theta_3 &=& +1 \quad\Longrightarrow\quad  
   \C_n,\s_n=0\;\text{for\ odd}\;n, 
 \nonumber\\
   \theta_2\theta_3 &=& -1 \quad\Longrightarrow\quad 
   \C_n,\s_n=0\;\text{for\ even}\;n.
 \eea     
In contrast to Eqs.~(\ref{19}) this is now true for all rapidities $Y$.
The corresponding Fourier expansions are listed in Table~\ref{T2}. Note
that according to Eq.~(\ref{19a}) the non-vanishing coefficients $G_n$,
$H_n$ and $I_n$ in Table~\ref{T2} are odd functions of rapidity and 
thus vanish at $Y\Eq2$.

\section{Fourier expansion of the radius parameters}
\label{sec4}

We will now combine the above implicit $\Phi$ dependence of the spatial
correlation tensor with the explicit $\Phi$ dependence shown in 
Eqs.~(\ref{5}). When studying the combination of symmetries I and II
for sources which are not invariant under longitudinal boosts, we 
must restrict our attention to midrapidity pairs with $\beta_l\Eq0$. 
For simplicity, we also set $\beta_l\Eq0$ in the boost-invariant case;
this means that we are studying the correlation radii in the 
longitudinally comoving system (LCMS) \cite{WH99}. General expressions 
for $\beta_l\ne0$ are easily obtained by boosting from the LCMS to a 
fixed longitudinal reference frame and can be found in 
Refs.~\cite{CSH95,CNH95}. For boost-invariant sources $R_{sl}^2\Eq0$
independent of rapidity \cite{CSH95,CNH95}; without boost invariance 
this is generally not even true at midrapidity (see Eqs.~(\ref{5}) 
and Table~\ref{T1}). As shown in Ref.~\cite{LHW00}, a non-zero value 
for $R_{sl}^2$ arises naturally if the longitudinal major axis of the 
source ellipsoid is tilted away from the beam direction; longitudinal 
boost invariance forbids such a tilt. 

Using the symbols introduced in Tables~\ref{T1} and \ref{T2} and
setting $\beta_l\Eq0$, Eqs.~(\ref{5}) simplify to
  \bea
    R_s^2 &=& A - B\cos(2\Phi) - C\sin(2\Phi),
  \nonumber\\
    R_o^2 &=& A + B\cos(2\Phi) + C\sin(2\Phi)
  \nonumber\\
          && - 2\beta_\perp (E\cos\Phi{+}F\sin\Phi)
             + \beta_\perp^2 D,
   \nonumber\\
    R_{os}^2 &=& C\cos(2\Phi) - B\sin(2\Phi) 
                 + \beta_\perp (E\sin\Phi{-}F\cos\Phi),
   \nonumber\\
    R_{l}^2 &=& J,
   \nonumber\\
    R_{ol}^2 &=& H\cos\Phi + I\sin\Phi - \beta_\perp G,
   \nonumber\\
    R_{sl}^2 &=& I\cos\Phi - H\sin\Phi .
    \label{22}
  \eea
Here $A,B,\dots,J$ are functions of $\Phi$ whose Fourier expansions are
given in Tables~\ref{T1} and \ref{T2}. For a boost-invariant source 
$G(\Phi),\,H(\Phi)$ and $I(\Phi)$ (i.e. $R_{ol}^2$ and $R_{sl}^2$) 
vanish at $\beta_l\Eq{Y}\Eq0$. A comparison of Tables~\ref{T1} and 
\ref{T2} shows that at $Y\Eq0$ all other $S_{\mu\nu}$ components 
entering Eqs.~(\ref{22}) have exactly the same Fourier expansion with 
and without longitudinal boost invariance. We may therefore investigate 
Eqs.~(\ref{22}) on the basis of the expansions listed in Table~\ref{T1} 
and recover the boost-invariant case later by simply setting 
$R_{ol}\Eq{R_{sl}}\Eq0$. 

In Ref.~\cite{LHW00} we studied the limit of vanishing implicit $\Phi$ 
dependence (i.e. $S_{\mu\nu}$ does not depend on $\Phi$). Table~\ref{T1}
shows that in this limit only the diagonal elements and $S_{13}$ (i.e. 
$A,\,B,\,D,\,H$ and $J$) are non-zero. As discussed in \cite{LHW00}, 
this limit requires only that space-momentum correlations (e.g. due 
to collective flow effects) are weak. A more general situation was 
analyzed in \cite{W97} where terms up to $n\Eq2$ were kept in the 
Fourier expansion of the HBT radii, but in that work the $\Phi$ 
dependences of the emission duration and time-space correlations 
were neglected relative to those of the spatial components $S_{ij}$.
We here remove both of these approximations.

Inserting the expansions in Table~\ref{T1} into (\ref{22}) and using
 \bea
 \label{23} 
   \cos n\Phi\,\cos m\Phi &=& 
   {\textstyle\2} \left[\cos(n{-}m)\Phi + \cos(n{+}m)\Phi\right],
 \nonumber\\
   \sin n\Phi\, \sin m\Phi &=& 
   {\textstyle\2} \left[\cos(n{-}m)\Phi - \cos(n{+}m)\Phi\right],
 \nonumber\\
   \cos n\Phi\,\sin m\Phi &=& 
   {\textstyle\2} \left[\sin(n{+}m)\Phi - \sin(n{-}m)\Phi\right],
 \eea
we see that at midrapidity the HBT radius parameters 
$R_{\alpha}^2(Y{=}0,\Kt,\Phi)$ have the following Fourier expansions:
 \bea
 \label{24}
   R_s^2 &=& R_{s,0}^2 + {\textstyle2\sum_{n=2,4,6,\dots}} 
   R_{s,n}^2\cos(n\Phi),
 \nonumber\\
   R_o^2 &=& R_{o,0}^2 + {\textstyle2\sum_{n=2,4,6,\dots}} 
   R_{o,n}^2\cos(n\Phi),
 \nonumber\\
   R_{os}^2 &=& \qquad\quad
   {\textstyle2\sum_{n=2,4,6,\dots}} R_{os,n}^2\sin(n\Phi),
 \nonumber\\
   R_l^2 &=& R_{l,0}^2 + {\textstyle2\sum_{n=2,4,6,\dots}}
   R_{l,n}^2\cos(n\Phi),
 \nonumber\\
   R_{ol}^2 &=& \qquad\quad
   {\textstyle2\sum_{n=1,3,5,\dots}} R_{ol,n}^2\cos(n\Phi),
 \nonumber\\
   R_{sl}^2 &=& \qquad\quad 
   {\textstyle2\sum_{n=1,3,5,\dots}} R_{sl,n}^2\sin(n\Phi).
 \eea
Due to the symmetries of the emission function, the HBT radius parameters 
are sums of either cosine or sine terms, involving either even or odd 
multiples of the emission angle $\Phi$, but no mixtures of different such 
terms. As a consequence, $R_{os}^2$ vanishes at both $\Phi\Eq0^\circ$ and 
$90^\circ$ (i.e. its leading contribution features a second order harmonic 
oscillation as a function of the emission angle). $R_{ol}^2$ and $R_{sl}^2$ 
in general exhibit leading first order harmonic oscillations \cite{LHW00}
with zeroes at $90^\circ$ and $0^\circ$, respectively. For a boost 
invariant source they vanish identically.

The Fourier coefficients $R_{\alpha,n}^2$ are functions of $\Kt$. We now
list them up to order $n\Eq2$. The $\Phi$-independent terms are given by
 \bea
 \label{25}
   R_{s,0}^2 &=& A_0-B_2-C_2,
 \nonumber\\
   R_{o,0}^2 &=& A_0+B_2+C_2 -2\beta_\perp(E_1+F_1)+\beta_\perp^2 D_0,
 \nonumber\\
   R_{l,0}^2 &=& J_0.
 \eea
The coefficients of the first order harmonics are
 \bea
 \label{26}
   R_{ol,1}^2 &=& {\textstyle\2}(H_0+H_2+I_2)-\beta_\perp G_1,
 \nonumber\\
   R_{sl,1}^2 &=& {\textstyle\2}(-H_0+H_2+I_2).
 \eea
For a boost-invariant source these vanish. The term ${\sim\,}H_0$ 
describes the tilt of the emission region relative to the beam axis 
which was discussed in \cite{LHW00}. The second order harmonic 
oscillations have amplitudes
 \bea
 \label{27}
   R_{s,2}^2 &=& A_2 -{\textstyle\2}(B_0+B_4+C_4),
 \nonumber\\
   R_{o,2}^2 &=& A_2 +{\textstyle\2}(B_0+B_4+C_4) 
 \nonumber\\
             && -\beta_\perp(E_1+E_3-F_1+F_3)+\beta_\perp^2 D_2,
 \nonumber\\
   R_{os,2}^2 &=& {\textstyle\2}(-B_0+B_4+C_4) 
                + \frac{\beta_\perp}{2}(E_1-E_3-F_1-F_3),
 \nonumber\\
   R_{l,2}^2 &=& J_2.
 \eea
If the emission duration $D\Eq\langle\tilde t^2\rangle$ is independent 
of emission angle ($D_{2,4,6,\dots}{\,\approx\,}0$) and all higher order 
harmonics $n{\,\geq\,}3$ of the spatial correlation tensor are small,
these amplitudes fulfill the approximate ``sum rule'' \cite{W97} 
 \begin{eqnarray}
 \label{28} 
   && R_{o,2}^2-R_{s,2}^2 + 2 R_{os,2}^2 =
 \nonumber\\ 
   && 2(B_4{+}C_4)-2\beta_\perp(E_3{+}F_3) +\beta_\perp^2 D_2.
      \approx 0,
 \end{eqnarray}
Note that the leading first order harmonics of $\langle \tilde t
\tilde x\rangle$ and $\langle \tilde t \tilde y\rangle$, which describe
how the transverse positions are correlated with time at freeze-out, 
cancel in this ``sum rule''. If the data satisfy this ``sum rule''
for all values of $\Kt$ resp. $\beta_\perp$, one may conclude (barring 
unlikely accidental cancellations among the terms) that $D_2$, 
$E_3$, $F_3$, $B_4$ and $C_4$ all vanish. In this case the azimuthal 
oscillation amplitudes of the transverse HBT radii reduce to 
 \bea
 \label{29}
   R_{s,2}^2 &=& A_2 -{\textstyle\2}B_0,
 \nonumber\\
   R_{o,2}^2 &=& A_2 +{\textstyle\2}B_0 -\beta_\perp(E_1{-}F_1),
 \nonumber\\
   R_{os,2}^2 &=& \quad -{\textstyle\2}B_0  
               + {\textstyle\2}\beta_\perp(E_1{-}F_1).
 \eea
The term ${\sim\,}(E_1{-}F_1)$ is the leading (first harmonic) 
contribution to the correlation $\langle(\tilde x{-}\tilde y)\tilde
 t\rangle$ between emission points and times. In a hydrodynamic model
this term reflects the geometric manifestation of elliptic flow,
namely that the freeze-out radius increases with time more rapidly
in $x$ than in $y$ direction \cite{KH02}. Since it comes with an 
explicit factor of $\beta_\perp$, one may be able to isolate it 
using the $\Kt$-dependence of the azimuthal oscillation amplitudes
(\ref{29}) at small $\Kt$.

Finally, in the absence of dynamical space-momentum correlations, all
implicit $\Phi$ dependences (i.e. all higher harmonics in Table~\ref{T1})
are expected to vanish, leading to the ``geometric relations'' \cite{W97}
 \bea
 \label{30}
   &&R_{s,0}^2 = A_0,
 \nonumber\\
   &&R_{o,0}^2 - R_{s,0}^2 = \beta_\perp^2 D_0,
 \nonumber\\
   &&R_{l,0}^2 = J_0,
 \nonumber\\
   &&R_{ol,1}^2 = - R_{sl,1}^2 = {\textstyle\2}H_0,
 \nonumber\\
   &&R_{o,2}^2 = - R_{s,2}^2 = - R_{os,2}^2 = {\textstyle\2}B_0.
 \eea
In this case all five non-vanishing components of the spatial correlation
tensor can be separated \cite{LHW00}.   

\section{Considerations for a finite symmetric window around $Y\Eq0$}
\label{sec5}

The results quoted so far were derived at midrapidity $Y\Eq0$ since, 
at least in the absence of longitudinal boost invariance, symmetry II 
can only there be used to constrain the azimuthal Fourier expansion 
of $S_{\mu\nu}$, by eliminating either even or odd terms in the sums 
over $n$ (see Eq.~(\ref{19})). In practice, statistical limitations 
render strict cuts on the pair rapidity $Y$ quite painful. It is 
therefore important to assess the necessary modifications if the 
data are collected in a finite size rapidity interval around $Y\Eq0$. 
We now prove the important result that, as long as the HBT radii are 
obtained from averaging over a {\em symmetric} rapidity interval 
around $Y\Eq0$, the general form (\ref{24}) of their Fourier expansions 
remains unchanged. On the other hand, equations (\ref{25}) -- (\ref{27}) 
receive additional contributions which, at leading order in the 
width $\Delta Y$ of the rapidity interval, grow quadratically as 
$(\Delta Y)^2$; this can be used to eliminate them by varying $\Delta Y$ 
and extrapolating to $\Delta Y\Eq0$. 

As noted in Eq.~(\ref{19a}), the point reflection symmetry (\ref{10}),
(\ref{11}) allows to classify the Fourier expansion coefficients of 
the spatial correlation tensor $S_{\mu\nu}$ as either even or odd 
functions of rapidity $Y$. The odd terms vanish at $Y\Eq0$, but do 
not do so any longer at $Y{\,\ne\,}0$. However, when calculating the 
HBT radii from $S_{\mu\nu}$ according to Eqs.~(\ref{5}) and averaging 
them over a finite, but symmetric rapidity interval around $Y\Eq0$, 
terms which are odd in $Y$ average to zero. Therefore, there are no 
new contributions to $R_s^2$, $R_o^2$ and $R_{os}^2$ in this case. 
$R_l^2$, $R_{ol}^2$ and $R_{sl}^2$, on the other hand, contain at 
$Y{\,\ne\,}0$ additional terms beyond those listed in Eqs.~(\ref{22}) 
which are multiplied by either one or two powers of $\beta_l$. When 
multiplying an odd Fourier coefficient by $\beta_l$, the result is 
even in $\beta_l$ (respectively $Y$) and does not average to zero 
across the rapidity interval $\Delta Y$. In fact, at leading order 
in $\Delta Y$, its average is $\sim\langle\beta_l^2\rangle$ which 
grows quadratically with $\Delta Y$.

Let us now look at how these extra terms modify the Fourier 
expansions given in the last three lines of Eq.~(\ref{24}). We begin 
with $R_l^2 = S_{33}-2\beta_l S_{03}+\beta_l^2S_{00}$ and average it
over the symmetric interval $\Delta Y$. Table~\ref{T1} tells us 
that the Fourier coefficients of $S_{33}$ with odd values of $n$ are
odd functions of $Y$ and thus average to zero. For $S_{03}$ the
coefficients with odd $n$ are even functions of $Y$, but since 
$S_{03}$ is multiplied by $\beta_l$, these odd $n$ terms again 
average to zero. The same is true for the last term where the
factor $\beta_l^2$ preserves the $Y$-reflection symmetries of 
the expansion coefficients. Altogether, the rapidity-averaged 
longitudinal radius $\langle R_l^2\rangle$ continues to have 
only even $n$ terms in its Fourier expansion, just as Eq.~(\ref{24}) 
states for $Y\Eq0$. In the same fashion one also shows that the 
rapidity-averaged radius parameters $\langle R_{ol}^2\rangle$ 
and $\langle R_{sl}^2\rangle$ continue to have the same Fourier 
expansions as in (\ref{24}). In other words, averaging the HBT radii 
over a finite, symmetric rapidity interval around $Y\Eq0$ preserves 
the general structure (\ref{24}) of their azimuthal Fourier expansions.

When expressing the Fourier components of the ra\-pi\-di\-ty-averaged HBT 
radii in terms of the harmonic coefficients of $S_{\mu\nu}$, new
terms arise, and Eqs.~(\ref{25})-(\ref{27}) are modified. We only 
list those equations whose structure changes:
 \bea
 \label{31}
   \langle R_{l,0}^2 \rangle &=& \langle J_0\rangle
      - 2\langle\beta_l G_0\rangle + \langle\beta_l^2 D_0\rangle, 
 \nonumber\\
   \langle R_{l,2}^2 \rangle &=& \langle J_2\rangle
      - 2\langle\beta_l G_2\rangle + \langle\beta_l^2 D_2\rangle, 
 \nonumber\\
   \langle R_{ol,1}^2 \rangle &=& 
     {\textstyle\2}\langle H_0{+}H_2{+}I_2 
                 - \beta_l(E_0{+}E_2{+}F_2)\rangle
\nonumber\\
   && -\beta_\perp \langle G_1{-}\beta_l D_1 \rangle,
 \nonumber\\
   \langle R_{sl,1}^2 \rangle &=& 
     {\textstyle\2}\langle{-}H_0{+}H_2{+}I_2 
                 - \beta_l({-}E_0{+}E_2{+}F_2)\rangle.\quad
 \eea
Here the angular brackets denote the average over the symmetric rapidity
interval $\Delta Y$. All terms involving one or two explicit factors 
$\beta_l$ vanish quadratically as $\Delta Y{\,\to\,}0$ in which limit
Eqs.~(\ref{25}) -- (\ref{27}) are recovered.  

\section{What if the sign of the impact parameter can not be determined?}
\label{sec5a}

If the orientation of the reaction plane is reconstructed from an even
Fourier component of the single particle distribution (e.g. from the
elliptic flow coefficient $v_2$ as is the case at RHIC), the direction
of the impact parameter vectoir $\bm{b}$ has a sign ambiguity, i.e. after
aligning events according to their reaction plane the event sample contains
equal contributions from collisions with impact parameters $\bm{b}$ and
$-\bm{b}$. This ambiguity does not exist if the event plane is 
reconstructed from the directed flow coefficient $v_1$ (as one does at 
the AGS and SPS) whose sign has a one-to-one correlation with the 
direction of $\bm{b}$ within the reaction plane.

If events with impact parameters $\bm{b}$ and $-\bm{b}$ are equally 
mixed, the effective source function is symmetric under the exchange 
$\bm{b}\to-\bm{b}$ which is equivalent to an azimuthal rotation by 
$180^\circ$:
 \bea
 \label{10a}
   \text{IIa}: && S(x,y,z,t;Y,\Kt,\Phi) 
 \nonumber\\
   &&\qquad\quad
     = S(-x,-y,z,t;Y,\Kt,\Phi+\pi).\quad
 \eea
For the spatial correlation tensor this implies
 \begin{equation}
 \label{11a} 
   \text{IIa}: S_{\mu\nu}(Y,\Kt,\Phi) = \theta_{2a}\,
   S_{\mu\nu}(Y,\Kt,\Phi{+}\pi),
 \end{equation}
with
 \begin{equation}
 \label{12a}
   \theta_2 = (-1)^{\delta_{\mu 1}+\delta_{\nu 1}+\delta_{\mu 2}
                   +\delta_{\nu 2}}.
 \end{equation}
One easily checks that the sign $\theta_{2a}$ is, in fact, equal to the
product $\theta_2\theta_3$ of the signs under symmetries I and III, as 
tabulated in Table~\ref{T2}. Correspondingly, the general form of the
Fourier expansions of $S_{\mu\nu}$ and of the HBT radii are exactly the 
same as those listed in and resulting from Table~\ref{T2} for a
longitudinally boost-invariant source. We see in particular that a sign 
ambiguity for the direction of the impact parameter automatically leads
to vanishing cross terms $R_{ol}^2$ and $R_{sl}^2$.  

\section{Corrections for binning and finite event plane resolution}
\label{sec6}

Experimentally, the two-pion correlation function is obtained as the
ratio of correlated pairs, $N(\bq,\bK)$, and uncorrelated (mixed event) 
pairs, $D(\bq,\bK)$.  In an azimuthally-sensitive analysis, one 
constructs these distributions for a given selection on emission 
angle $\Phi$.  However, finite binning in $\Phi$ and uncertainty in 
the experimental estimation of the reaction plane tend to dampen the 
azimuthal dependencies of the observed (``raw'') distributions 
$N_{\rm exp}(\bq,\bK)$ and $D_{\rm exp}(\bq,\bK)$. In this Section we 
present a model-independent procedure to correct for these effects.

Since the reaction plane is reconstructed event-by-event from the 
anisotropies of the single particle momentum distribution 
\cite{O93,VZ96,PV98}, its orientation is only known with a finite 
statistical accuracy controlled by the number of particles used in 
the reconstruction process. Correspondingly, in a statistical 
average over the event sample the true reaction plane angle $\psi_R$
is distributed around the reconstructed one $\psi_m$ by a probability
distribution (see Eq.~(9) in \cite{PV98})
 \bea
 \label{32}
   &&p(\psi_m{-}\psi_R) \equiv 
   \frac{dP}{d\left(m(\psi_m{-}\psi_R)\right)} =
   \int \frac{v'_m\,dv'_m}{2\pi\sigma^2}
 \nonumber\\
   && \times\exp\left[-\frac{v_m^2{+}{v'_m}^2{-}2
                    v_m v'_m\cos\left(m(\psi_m{-}\psi_R)\right)}
                   {2\sigma^2}\right],\quad
 \eea
with width $\sigma^2=\langle w^2\rangle/(2M\langle w\rangle^2)$ (where
$M$ is the number of particles per event and $w$ is an arbitrary weight 
function (e.g. $w\Eq1$ or $w{\Eq}p_\perp$) used in the analysis). $m$ 
denotes the order of the Fourier component of the single-particle 
spectrum used to extract the reaction plane, and $v_m$ is the 
corresponding Fourier coefficient; the cases $m\Eq1$ (directed flow)
and $m\Eq2$ (elliptic flow) are relevant in practice. Correspondingly, 
a measurement of the two-particle distributions $N(\bq,\bK)$ and 
$D(\bq,\bK)$ at fixed emission angle $\Phi\equiv\Phi{-}\psi_m$ relative 
to the reconstructed event plane corresponds to an {\em average} of 
the real two-particle distributions over a range of emission angles 
$\Phi{-}\psi_R$ relative to the true reaction plane, where the average 
is taken with the distribution (\ref{32}). The averaging reduces the 
azimuthal dependence of the correlation function (and of the HBT radii 
extracted from it) and must be corrected for, before comparing to models.

An additional smearing which goes in the same direction arises from the
binning of the data in $\Phi$. By summing the data over all emission
angles $\phi$ within an interval of width $\Delta$ centered at $\Phi$,
one effectively performs an additional smearing of $N(\bq,\bK)$ and 
$D(\bq,\bK)$ over the azimuthal emission angle with the distribution
 \begin{equation}
 \label{33}
   f_\Delta(\phi{-}\Phi) = \frac{1}{\Delta} 
   \theta\left(\phi{-}\Phi+\half\Delta\right) 
   \theta\left(\half\Delta{-}\phi{+}\Phi\right).
 \end{equation}
The two effects can be combined by folding the distributions $p$ and 
$f_\Delta$ and averaging the true correlated and mixed pair distributions 
with 
 \begin{equation}
 \label{34}
   H_\Delta(\phi) = \int_{-\pi}^\pi d\psi\ p(\psi)\ f_\Delta(\phi-\psi)
   = \frac{1}{\Delta} \int_{-\Delta/2}^{\Delta/2} \!\!\!\! 
     d\theta\ p(\phi-\theta).
 \end{equation}

The above azimuthal averaging affects the numerator $N$ and denominator 
$D$ separately. Suppressing the dependence on $\Kt$ and $Y$ for clarity, 
the measured angular dependence of the correlated pairs relative to the 
reconstructed reaction plane $\psi_m$, $N_{\rm exp}(\bq,\Phi{-}\psi_m)$,
is related to their true angular dependence relative to the real
reaction plane $\psi_R$, $N(\bq,\Phi{-}\psi_R)$, by
 \bea
 \label{35}
   N_{\rm exp}(\bq,\Phi{-}\psi_m) &=& \int_{-\pi}^\pi d\phi\ 
   N_\Delta(\bq,\phi{-}\psi_R)
 \nonumber\\
   &\times& p\bigl((\Phi{-}\psi_m)-(\phi{-}\psi_R)\bigr), 
 \eea
where 
 \begin{equation}
 \label{36}
   N_\Delta(\bq,\phi{-}\psi_R) = 
   \int_{\phi-\psi_R-\Delta/2}^{\phi-\psi_R+\Delta/2}
   N(\bq,\theta)\,d\theta
 \end{equation}
denotes the effect of summing the data in angular bins of width $\Delta$.
An analogous pair of equations holds for the measured and true 
uncorrelated pairs in the denominator, $D_{\rm exp}(\bq,\Phi{-}\psi_m)$ and
$D(\bq,\phi{-}\psi_R)$. 

The task at hand is to extract the true angular dependence on 
$\Phi{-}\psi_R$ from the measured dependence on $\Phi_j{-}\psi_m$
where $j$ labels the angular bins centered at angles $\Phi_j$ 
relative to the reconstructed reaction plane. To this end we Fourier
decompose the measured quantities $N$ and $D$ for each value of $\bq$. 
For example
\begin{widetext}
 \bea 
 \label{37}
   && N_{\rm exp}(\bq,\Phi{-}\psi_m) = N^{\rm exp}_0(\bq) +
      2\sum_{n=1}^{n_{\rm bin}} 
        \bigl[N^{\rm exp}_{c,n}(\bq)\cos(n(\Phi{-}\psi_m))
            + N^{\rm exp}_{s,n}(\bq)\sin(n(\Phi{-}\psi_m))\bigr],
 \nonumber\\
   && N^{\rm exp}_{c,n}(\bq) \equiv 
    \langle N_{\rm exp}(\bq,\Phi) \cos(n \Phi) \rangle   
    = \frac{1}{n_{\rm bin}} \sum_{j=1}^{n_{\rm bin}} 
          N_{\rm exp}(\bq,\Phi_j) \cos(n \Phi_j), 
 \nonumber\\
    && N^{\rm exp}_{s,n}(\bq) \equiv
    \langle N_{\rm exp}(\bq,\Phi) \sin(n\Phi) \rangle 
    = \frac{1}{n_{\rm bin}} \sum_{j=1}^{n_{\rm bin}}
        N_{\rm exp}(\bq,\Phi_j) \sin(n\Phi_j),
 \eea
where $n_{\rm bin}$ denotes the number of angular bins (for finite 
$n_{\rm bin}$ only Fourier components with $n{\,\leq\,}n_{\rm bin}$ are
meaningful). We further imagine doing the same for the 
corresponding true and binned quantities corrected for event-plane 
resolution:
 \bea
 \label{38}
  && N(\bq,\Phi{-}\psi_R) = N_0(\bq) + 2\sum_{n=1}^{n_{\rm bin}}
    \bigl[N_{c,n}(\bq)\cos(n(\Phi{-}\psi_R)) 
        + N_{s,n}(\bq)\sin(n(\Phi{-}\psi_R))\bigr],
 \nonumber\\[1ex]
  && N_\Delta(\bq,\theta) = N_0^{\Delta}(\bq) + 2\sum_{n=1}^{n_{\rm bin}}
   \bigl[N_{c,n}^{\Delta}(\bq)\cos(n\theta) 
       + N_{s,n}^{\Delta}(\bq)\sin(n\theta)\bigr].
 \eea
\end{widetext}
Analogous expressions hold for the mixed pairs in the denominator $D$.
Inserting the Fourier expansions (\ref{38}) into Eqs.~(\ref{35}) and
(\ref{36}), using that the distributions $p$ and $f_\Delta$ are even
functions of their arguments, and comparing the result with Eq.~(\ref{37}) 
one easily finds for all $n$ and both series of coefficients 
($\alpha\Eq{c}$ or $s$)
 \bea
 \label{39}
   N_{\alpha,n}^{\Delta}(\bq) &=& N_{\alpha,n}(\bq)\,
   \frac{\sin(n\Delta/2)}{n\Delta/2},
 \nonumber\\
   N^{\rm exp}_{\alpha,n}(\bq) &=& N_{\alpha,n}^{\Delta}(\bq) \ 
   \bigl\langle\cos\left(n(\psi_m{-}\psi_R)\right)\bigr\rangle_p.
 \eea
The factors $\langle\cos\left(n(\psi_m{-}\psi_R)\right)\rangle_p$, 
arising from an average over the event plane distribution (\ref{32}), 
are the well-known correction factors for event-plane resolution 
arising in the process of extracting the anisotropic flow coefficients 
$v_n$ from the single particle spectrum \cite{O93,VZ96,PV98}.

With the results (\ref{39}), the numerator $N$ and denominator $D$ of 
the correlation function at each measured angle $\Phi_j$ and relative 
momentum $\bq$ can now be easily corrected for the effects of angular 
binning and finite event plane resolution (setting again $\psi_m\Eq0$):
\begin{widetext}
 \begin{equation}
 \label{40}
   N(\bq,\Phi_j) = N_{\rm exp}(\bq,\Phi_j) +
   2\sum_{n=1}^{n_{\rm bin}} \zeta_{n,m}(\Delta)
     \bigl[N^{\rm exp}_{c,n}(\bq)\cos(n\Phi_j) 
         + N^{\rm exp}_{s,n}(\bq)\sin(n\Phi_j)\bigr],\qquad
 \end{equation}
\end{widetext}
with correction parameters $\zeta_{n,m}(\Delta)$ given by the simple
expression
 \begin{equation}
 \label{41}
   \zeta_{n,m}(\Delta)=\frac{n\Delta/2}
   {\sin(n\Delta/2)\langle\cos(n(\psi_m{-}\psi_R))\rangle_p} - 1.
 \end{equation}
A similar equation holds for the uncorrelated pairs in the denominator 
$D$. Since the right hand side of Eq.~(\ref{40}) involves only 
experimentally known quantities, the correction algorithm is model 
independent. The sums over $n$ go over all allowed values; if $m$ is 
even (i.e. the sign of the impact parameter is not known), both $N$ 
and $D$ are symmetric under azimuthal rotations by $180^\circ$ and 
only even values of $n$ are summed over. Contrary to the HBT radius 
parameters or to single-particle flow measures~\cite{PV98}, $N$ and $D$
have no unique symmetry under $\Phi{\,\to\,}-\Phi$; 
thus in general both sine and cosine terms contribute to the sum in 
Eq.~(\ref{40}).

After applying the algorithm (\ref{40}) to the data, the ratio 
$C(\bq,\bK){\Eq}N(\bq,\bK)/D(\bq,\bK)$ gives the corrected 
two-particle correlation function from which all angular binning and 
event plane resultion effects have been removed. The true emission
angle dependence of the HBT radius parameters can thus be directly 
extracted from a Gaussian fit with Eq.~(\ref{3}) to this function 
$C(\bq,\bK)$.

\section{Summary}
\label{sec7}

Equations (\ref{24}) give the most general Fourier expansions for the 
HBT radius parameters at midrapidity which are consistent with the 
symmetries of the source in {\em non-central} collisions between equal 
mass spherical nuclei. For full-overlap central collisions between 
{\em deformed nuclei} (e.g. U+U) and for longitudinally 
boost-invariant sources they also apply at $Y{\,\ne\,}0$. The structure 
of these expansions is preserved if the data are averaged over a 
symmetric finite rapidity interval around $Y\Eq0$. They provide 
a basis for fitting the azimuthal emission angle dependence of 
experimentally determined correlation functions. A model-independent 
correction of the measured two-pion correlation function for event 
plane resolution and angular binning effects is possible and given
in Section~\ref{sec6}. Equations~(\ref{25})--(\ref{27}), (\ref{30}) 
and (\ref{31}) relate the oscillation amplitudes extracted from the 
thus corrected correlation function to the leading harmonic 
coefficients of the spatial correlation tensor and allow to constrain 
models for the emission function using azimuthally sensitive HBT data. 
Under favorable conditions spatial and temporal aspects of the emission 
function can be separated.

\section*{Acknowledgments}

We gratefully acknowledge fruitful discussions with P. Kolb and 
B. Tom\'a\v{s}ik. This work was supported in part by the U.S. Department 
of Energy under Contract No. DE-FG02-01ER41190 and by the National 
Science Foundation under Grants No. PHY99-07949 and PHY00-99476. 
U.H. acknowledges the warm hospitality of the Kavli Institute for Theoretical 
Physics at the University of Santa Barbara during the Workshop Program 
``QCD in the RHIC Era''.

\bigskip


\end{document}